\documentclass[reprint,longbibliography,amsmath,amssymb,aps,nofootinbib]{revtex4-1}
\usepackage{multirow} 
\usepackage{graphicx}
\usepackage{dcolumn}
\usepackage{bm}
\usepackage{amssymb,amsmath}
\usepackage{tabularx}
\usepackage{subcaption}
\usepackage{flushend}
\usepackage{lipsum}
\usepackage{float}
\usepackage{amsmath,multirow,bigdelim}
\usepackage{contour}
\usepackage{array}
\usepackage[bottom]{footmisc}
\usepackage{soul}
\usepackage{hyperref}
\newcolumntype{P}[1]{>{\centering\arraybackslash}p{#1}}
\usepackage{rotating} 

\usepackage{amssymb}
\usepackage{pifont}
\newcommand{\cmark}{\ding{51}}%
\newcommand{\xmark}{\ding{55}}%

\newcommand{\beq}{\begin{equation}}
\newcommand{\eeq}{\end{equation}}
\newcommand{\bea}{\begin{eqnarray}}
\newcommand{\eea}{\end{eqnarray}}
\newcommand{\ba}{\begin{array}} 
\newcommand{\ea}{\end{array}}

\newcommand{\ii}{\boldsymbol{i}}

\newcommand\gev{\,\mathrm{GeV}}
\newcommand\tev{\,\mathrm{TeV}}
\newcommand\eg{{\it e.g.}}
\newcommand\ie{{\it i.e.}}
\newcommand{\gsim}{\lower.7ex\hbox{$\;\stackrel{\textstyle>}{\sim}\;$}}
\newcommand{\lsim}{\lower.7ex\hbox{$\;\stackrel{\textstyle<}{\sim}\;$}}

\newcommand\rdstar{R_{D^*}}
\newcommand\rd{R_{D^{(\!*\!)}}}
\newcommand\rdrdstar{\rd}
\newcommand\mlq{M_{\textrm LQ}}
\newcommand{\yud}{Y_u^\dagger}
\newcommand{\ydd}{Y_d^\dagger}

\definecolor{brown}{rgb}{.7,.2,.2}
\definecolor{violet}{rgb}{.6,.3,.8}

\begin{document}

\title{On the Minimal Flavor Violating Leptoquark Explanation of the $\rd$ Anomaly}

\author{Saurabh Bansal}
\author{Rodolfo M.~Capdevilla}
\author{Christopher Kolda}

\affiliation{Department of Physics, University of Notre Dame, 225 Nieuwland Hall, Notre Dame, Indiana 46556, USA}

\begin{abstract}
There has been persistent disagreement  between the Standard Model (SM) prediction and experimental measurements of $R_{D^{(\!*\!)}}=\mathcal{B}(\bar B \rightarrow D^{(*)} \tau \bar\nu_\tau)/\mathcal{B}(\bar B \rightarrow D^{(*)} l \bar\nu_l)$ $(l=e,\mu)$. This anomaly may be addressed by introducing interactions beyond the Standard Model involving new states, such as leptoquarks. Since the processes involved are quark flavor changing, any new states would need to couple to at least two different generations of quarks, requiring a non-trivial flavor structure in the quark sector while avoiding stringent constraints from flavor-changing neutral current processes. In this work, we look at scalar leptoquarks as a possible solution for the $R_{D^{(\!*\!)}}$ anomaly under the assumption of {\it minimal flavor violation} (MFV). We investigate all possible representations for the leptoquarks under the SM quark flavor symmetry group, consistent with asymptotic freedom. We consider constraints on their parameter space from self-consistency of the MFV scenario, perturbativity, the FCNC decay $b\to s\bar\nu\nu$ and precision electroweak observables. We find that none of the scalar leptoquarks can explain the $R_{D^{(\!*\!)}}$ anomaly while simultaneously avoiding all constraints within this scenario. Thus scalar leptoquarks with MFV-generated quark couplings do not work as a solution to the $R_{D^{(\!*\!)}}$ anomaly.
\end{abstract}

\maketitle

\section{\label{section1}Introduction}

Precision measurements in the quark flavor sector are known to be fertile ground for probing physics at energy scales much higher than those that can be accessed in direct production experiments. In particular, flavor-changing neutral currents (FCNCs) can potentially probe physics at scales above $100\tev$ due to their GIM- and loop-suppressed amplitudes within the Standard Model (SM). In such cases, new physics that generates FCNC amplitudes at tree level or without GIM suppression could potentially dominate the SM contribution, making them a particularly strong probe of new physics. For this same reason, the ability of charged-current (CC) weak interactions to probe new physics is sharply limited, as the SM contributions are unsuppressed and new physics would have to show itself by (presumably small) interference effects with the SM amplitudes. Given the plethora of precision data on CC weak interactions, room for discovering new physics in this channel seems quite limited.

Nonetheless, over the last decade a growing set of measurements on the CC decays $B\to D\tau\nu$ and $B\to D^*\tau\nu$ have consistently shown an excess compared to the decays $B\to D^{{\scriptscriptstyle(}*{\scriptscriptstyle )}}\ell\nu$ where $\ell=e,\mu$. More specifically, one defines two observables:
\beq
R_D = \frac{\mbox{B}(\bar B\to D\tau\bar\nu)}{\mbox{B}(\bar B\to D\ell\bar\nu)}, \quad\quad 
\rdstar = \frac{\mbox{B}(\bar B\to D^*\tau\bar\nu)}{\mbox{B}(\bar B\to D^*\ell\bar\nu)}
\eeq
where $\ell=e,\mu$. (The data is consistent with $e-\mu$ universality in these decays, and thus one typically combines both final states in defining the ratio.)  The ratios are relatively insensitive to the uncertainties in the hadronic matrix elements; as a result, the SM predictions \cite{Aoki:2016frl,Bigi:2016mdz,Na:2015kha,Lattice:2015rga,Fajfer:2012vx,Bernlochner:2017jka,Bigi:2017jbd,Jaiswal:2017rve} are known at the 1-2\% level:
$$ R_D(\mathrm{pred}) =  0.299\pm 0.003, \quad \rdstar(\mathrm{pred}) = 0.258\pm 0.005. $$
To date, BaBar and Belle have both presented data on $R_D$ that, when combined, falls $2.3\sigma$ above SM expectations. For $\rdstar$, BaBar, Belle and LHCb have all presented data, which combine to be $3.0\sigma$ above SM expectations. Combining all current data \cite{Hirose:2016wfn,Sato:2016svk,Huschle:2015rga,Aaij:2015yra,Lees:2013uzd,Lees:2012xj,Aaij:2017uff,Aaij:2017deq} yields:
$$ R_D(\mathrm{exp}) = 0.407\pm 0.046, \quad \rdstar(\mathrm{exp}) = 0.306 \pm 0.015. $$
Taken together, these data appear to argue for new physics in CC interactions.

At the parton level, $R_D$ and $\rdstar$ can be reinterpreted in terms of the quark level processes $b\to c\tau\bar{\nu}_\tau$ and $b\to c\ell\bar{\nu}_\ell$. Thus, new physics that could affect this process must couple to second and third generation quarks and, must differentiate between the third generation of leptons and the first two. A strong candidate for such a state would be a charged Higgs boson.
Refs. \cite{Lees:2012xj,Tanaka:2012nw,Celis:2012dk} considered the interpretation of the data in terms of the type-II two-Higgs doublet model (2HDM) and found it to be inconsistent with the data. On the other hand, besides a plethora of model-independent analysis of the present anomaly \cite{Dutta:2018jxz,Alok:2018uft,Dutta:2017wpq,Alok:2017qsi,Ivanov:2017mrj,Bhattacharya:2016zcw,Ivanov:2016qtw,Ivanov:2015tru,Calibbi:2015kma,Alonso:2015sja,Bhattacharya:2014wla,Dutta:2013qaa,Goldberger:1999yh}, one can find in the literature potential explanations in terms of $W'$ vector bosons \cite{Boucenna:2016qad,Greljo:2015mma,He:2012zp}, composite states \cite{Marzocca:2018wcf,Fajfer:2012jt}, and Frogatt-Nielson-type models~\cite{Hiller:2016kry}. 
Alternatively, a number of authors have considered leptoquarks as a potential source for the anomaly \cite{Becirevic:2018afm,Blanke:2018sro,Monteux:2018ufc,Buttazzo:2017ixm,Altmannshofer:2017poe,Crivellin:2017zlb,Becirevic:2016yqi,Das:2016vkr,Barbieri:2015yvd,Fajfer:2015ycq,Bauer:2015knc,Freytsis:2015qca,Angelescu:2018tyl,Duraisamy:2014sna,Sakaki:2013bfa,Dorsner:2013tla,Hu:2018lmk}. This is the avenue that we will be considering in this paper as well.

As a candidate for new physics, leptoquarks are highly motivated, appearing naturally in any theory that unifies quarks and leptons into common multiplets or in $R$-parity violating models of supersymmetry. But leptoquarks present a number of problems as well. For a leptoquark coupling to first (second) generation quarks and leptons, direct production bounds at the LHC force the leptoquark mass above roughly $1.10 \tev \,(1.05 \tev)$ 
assuming it decays $100\%$ of the time to a charged lepton and a quark \cite{Aaboud:2016qeg,Khachatryan:2015vaa}; indirect searches for leptoquarks can impose even stronger constraints on their parameter space~\cite{Raj:2016aky,Dorsner:2016wpm,Bansal:2018eha}. In addition, leptoquark Lagrangians tend to violate both Baryon $(B)$ and Lepton $(L)$ numbers, leading to rapid proton decay. In this work, we will assume $B$ violating couplings to be zero in order to retain the stability of matter. And most importantly (for our purpose), unless one enforces some special flavor structure on the couplings of the leptoquark, large new contributions to $K^0-\bar K^0$ and $D^0-\bar D^0$ mixing force the mass of any such leptoquark to be larger than $O(100-1000\tev)$.

There are, however, special flavor structures that mitigate against new FCNC contributions, namely those implied by {\it minimal flavor violation} (MFV)~\cite{DAmbrosio:2002vsn,Nikolidakis:2007fc,Davidson:2010uu,Zhang:2018nmy}. MFV assumes that the approximate $[SU(3)]^3$ flavor symmetries present in the quark sector\footnote{Even though one can also realize MFV in the lepton sector, in this paper we will focus on MFV in the quark sector alone.} of the SM are broken only by the Yukawa interactions. MFV, in effect, ensures that the couplings of quarks to both Higgs and leptoquarks have (approximately) the same alignment. In so doing, moving to the quark mass basis generates couplings to the leptoquark that are nearly diagonal in generation space, with corrections proportional to CKM elements. It should be emphasized that MFV is a structure assumed and imposed on the leptoquark couplings that prevents the leptoquark from generating large FCNCs, but it is not itself a mechanism for generating such structures. This paper makes no attempt to provide an ultraviolet completion of the Lagrangians considered here, but once imposed it is technically natural. In addition, we will choose by fiat that the leptoquarks studied here couple only to the tau lepton and not the other charged leptons, as any explanation of the $R_{D^{(*)}}$ anomaly requires a clear violation of lepton universality.

One of the predictions from imposing an MFV structure on the leptoquark couplings is that the leptoquarks occur in multiplets of one or more of the $SU(3)$ quark flavor groups. The leptoquark capable of coupling both to $b$- and $c$-quarks will then necessarily have couplings to other quarks, enhanced or suppressed by CKM elements. One such scenario was originally studied in Ref.~\cite{Freytsis:2015qca}. We will revisit this model including other minimally flavor violating leptoquarks, and consider a set of constraints imposed by rare meson decays and precision electroweak measurements at LEP and SLC. The importance of electroweak data in the context of flavor anomalies, including $R_{D^{(*)}}$ anomaly, has been pointed out in Refs.~\cite{Feruglio:2016gvd,Feruglio:2017rjo,ColuccioLeskow:2016dox}.

More specifically, in this paper, we will examine the scalar leptoquarks in section II, and find that only two leptoquarks can provide an explanation of the $\rd$ anomaly. On imposing MFV in section III, we find that there are a total of twelve cases for these two leptoquarks that can possibly explain the anomaly. These cases refer to different transformation properties of the leptoquarks under the $SU(3)$ quark flavor groups. In section IV, we calculate the constraints on these leptoquarks from low energy measurements and find that there are two cases that can avoid these constraints and still explain the anomaly. We will then turn to the precision electroweak constraints, finding that
these exclude the remaining two cases, so no scalar leptoquark in the MFV framework can explain the $R_{D^{(*)}}$ anomaly. In section V, we present our conclusions.

\section{\label{section2}Operator Analysis}

Before specifying our leptoquark model, it is helpful to examine solutions to the $R_{D^{(*)}}$ anomaly within an effective Lagrangian approach. Here we update the fits found in Ref.~\cite{Freytsis:2015qca} with the most recent world averages from the Heavy Flavor Averaging Group (HFLAV)\footnote{\url{https://hflav.web.cern.ch/}}. The pieces of the effective Hamiltonian contributing to $b\to c\tau\bar{\nu}_\tau$ can be written as:
$$
\mathcal{H}_{eff} \supset 2 \sqrt{2} G_F V_{cb} \mathcal{O}_{V_L} + \frac{1}{\Lambda^2} \sum_i C_i \mathcal{O}_i
$$
where $\mathcal{O}_{V_L}$ is the SM operator, $(\bar{c}\gamma_\mu P_L b)(\bar{\tau} \gamma^\mu P_L \nu)$, and $\mathcal{O}_i$ ($C_i$) are the dimension-6 Wilson operators (and their coefficients). The complete list of dimension-6 operators that can contribute to $\bar{B} \rightarrow D^{(*)} \tau \bar{\nu}$ is given in Ref.~\cite{Freytsis:2015qca}. Of these operators, three can be mediated by scalar leptoquarks. These three operators, denoted $\mathcal{O}'_{S_L}$, $\mathcal{O}''_{S_L}$ and $\mathcal{O}''_{S_R}$, are listed in Table~\ref{tab:opTable} along with the SM quantum numbers of the corresponding scalar leptoquark(s). Of these three, the operator ${\cal O}_{S_R}''$ is identical to the $V-A$ operator of the SM (up to a factor of 2) after Fierzing, while the operator ${\cal O}_{S_L}''$ becomes a combination of scalar and tensor operators.

\begin{table}[t]
	\begin{ruledtabular}
		{\renewcommand{\arraystretch}{1.25}
			\begin{tabular}{c c c}
				\textrm{Name}&
				\textrm{Operator}&
				\textrm{Leptoquark}\\
				\hline 
				\rule{0pt}{3ex} 
				$\mathcal{O}'_{S_L}$ & $(\bar{\tau} P_L b)(\bar{c} P_L \nu)$ & $R_2~(3, 2, 7/2)$\\
				$~\mathcal{O}''_{S_L}$ & $(\bar{\tau} P_L c^c)(\bar{b}^c P_L \nu)$ &   $S_1~(\bar{3}, 1, 1/3)$ \\
				\multirow{2}{*}{$~\mathcal{O}''_{S_R}$}&\multirow{2}{*}{$(\bar{\tau} P_R c^c)(\bar{b}^c P_L \nu)$} &\rdelim\{{2}{7pt}$S_1~(\bar{3}, 1, 1/3)$\\
				&&$~~~S_3~(\bar{3}, 3, 1/3)$
				\rule{0pt}{3ex} 
			\end{tabular}
		}
	\end{ruledtabular}
	\caption{\label{tab:opTable}%
Dimension-6 operators contributing to $\bar{B} \rightarrow D^{(*)} \tau \bar{\nu}$, along with the scalar leptoquark(s) that can generate said operators. Next to each leptoquark are their charges under the SM group $SU(3)_c\times SU(2)_L\times U(1)_Y$.}
\end{table}

\begin{figure}[tb]
	\includegraphics[width=0.95\linewidth]{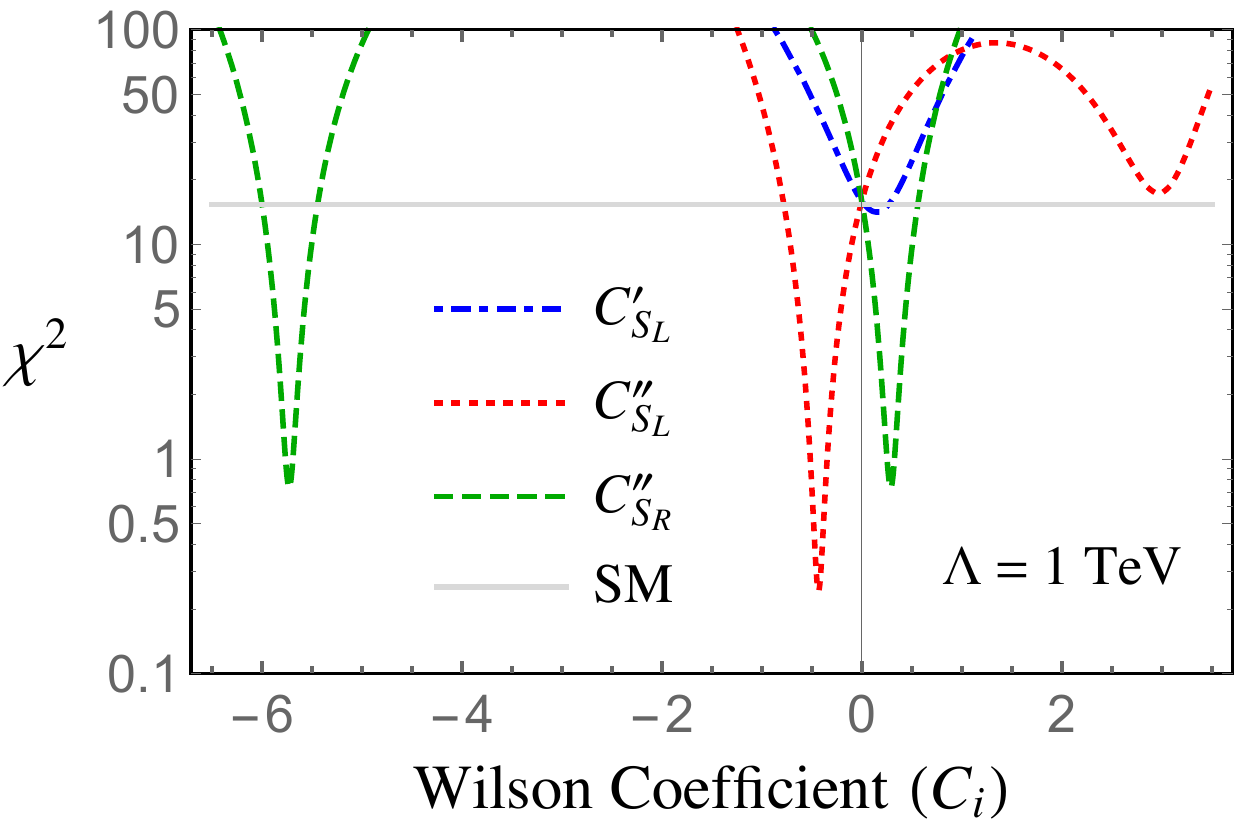}
	\caption{\label{fig:OperatorChiSq}$\chi^2$ values from a fit to the $\rd$ data, as a function of Wilson coefficients ($C_i$) for the operators generated by leptoquark exchange.}
\end{figure}

A $\chi^2$ analysis can be done for each operator separately, using as inputs the current combined best fit values for $R_D$ and $\rdstar$ given in section~\ref{section1}. The results of such a fit are shown in Fig.~\ref{fig:OperatorChiSq}. Here the scale $\Lambda$ is set to $1\tev$, and the value of $\chi^2$ is plotted as a function of the corresponding Wilson coefficient. One sees immediately that the operator $\mathcal{O}_{S_L}''$ or $\mathcal{O}_{S_R}''$ can provide a significantly improved fit to the $\rdrdstar$ data as compared to the SM (which is denoted with the solid line at $\chi^2\simeq 15$). On the other hand, the operator $\mathcal{O}_{S_L}'$ provides a fit to the data that is only minimally better than the SM in terms of total $\chi^2$, at the cost of an additional degree of freedom. We therefore judge this operator as providing a poor explanation for the anomaly and do not consider it further.

Thus, there are only two scalar leptoquarks that we need to consider further: the $S_1$, an $SU(2)_L$ singlet with hypercharge of $1/3$; and $S_3$, an $SU(2)_L$ triplet with hypercharge also of $1/3$. We will write the relevant pieces of their Lagrangians in the next section.

For both of the operators in which a good fit to the data is obtained, one finds two solutions with minimum $\chi^2$. For the operator $\mathcal{O}_{S_L}''$, we find: 
\beq
C''_{S_L} = -0.428\pm 0.096 
\label{CSLpp}
\eeq
while for $\mathcal{O}_{S_R}''$ we find:
\beq
C''_{S_R} = 0.293\pm 0.074  \,\,\,\,\, \mbox{ or }\,\, -5.72\pm 0.07
\label{CSRpp}
\eeq
for $\Lambda=1\tev$. These values are indicative of the lengths to which one must go in order to solve the $\rd$ anomaly by the exchange of new particles. If we identify the coefficients $C''$ with $\lambda^2/\mlq^2$, with $\lambda$ representing the coupling of the leptoquark to the fermions, and $\mlq$ its mass, then for $\lambda\sim O(g)$ one finds $\mlq\sim300\gev$ to $1.3\tev$, which are extremely low scales for new physics. But we should not be surprised to discover that such light (or, alternatively, strongly coupled) states are needed in order to modify charged current processes at the experimentally observed level. As we will soon see, the situation is appreciably worse once one imposes a realistic flavor structure on the leptoquark couplings to fermions.

\section{\label{setion3}Leptoquark Interactions and Flavor Structure}

The most general Lagrangians for $S_1$ and $S_3$ leptoquarks that preserve both $B$ and $L$ can be written as:
\beq
{\cal L}_{S_1} = S_1 \left\{ \lambda_{ij} \bar Q^c_{i}\, i\tau_2\, L_{j} + \tilde\lambda_{ij} \bar u^c_{i} e_{j} + h.c.\right\}
\label{LagS1}
\eeq
and 
\beq
{\cal L}_{S_3} = S_3^a \, \lambda_{ij} \bar Q^c_{i}\, i\tau_2 \tau_a\, L_{j} + h.c.
\label{LagS3}
\eeq
In the expressions above, $i,j$ are generation indices, $Q_i$ and $L_j$ are the left-chiral quark and lepton doublets, $u_i$ and $e_j$ are the right-chiral quark and lepton singlets, and $\tau_a$ $(a = 1, 2, 3)$ are the Pauli matrices. 

Because of the form of its interactions, the leptoquark $S_3$ can only generate $\mathcal{O}_{S_R}''$. However, $S_1$ can generate both operators. In particular, the first term in the Lagrangian of $S_1$ can generate the operator $\mathcal{O}_{S_R}''$, and the combination of both terms can lead to $\mathcal{O}_{S_L}''$.  Specifically:
$$C_{S_R}'' \propto |\lambda_{1L}|^2, \quad C_{S_L}''\propto \lambda_{1L}\lambda_{1R}.$$
Thus, if one wants to study the operator ${\cal O}_{S_L}''$, one must by necessity allow ${\cal O}_{S_R}''$ of roughly the same order of magnitude. This means that the single parameter fit to $C_{S_L}''$ in the last section is incomplete. Instead one must do a two-parameter fit simultaneously to $C_{S_L}''$ and $C_{S_R}''$. Such a fit yields regions of good fit to the $\rd$ data:\\ ~~ \\
$\mbox{Region 1}: C_{S_R}''= 0.69 \pm 0.15, \quad C_{S_L}'' = 2.58 \pm 0.23$ \\~~\\
and \\~~\\
$\mbox{Region 2}: C_{S_R}''= -6.13\pm 0.16, \quad C_{S_L}'' = -2.58 \pm 0.23.$\\~~\\
There are two additional regions in which the fit is almost entirely due to $C_{S_R}''$ with $C_{S_L}''\simeq 0$; these two regions are essentially identical to the single parameter fit of $C_{S_R}''$ from the previous section. Importantly, of the regions found by the two-parameter fit, the Lagrangian for $S_1$ can never provide a solution in Region 2, since the sign of the contributions of the $S_1$ leptoquark to $C_{S_R}''$ is always positive. In addition, both regions can be ruled out in a model-independent way by the lifetime of the $B_c$ meson, which receives a large contribution from the $C_{S_L}''$ operator~\cite{Li:2016vvp,Huang:2018nnq,Angelescu:2018tyl,Celis:2016azn}. We will not consider further any leptoquark models that require large $C_{S_L}''$ contributions in order to explain the $\rd$ anomaly.  There are also potentially large contributions from the $C_{S_R}''$ operator to the $B_c$ lifetime, but we find that the regions of $C_{S_R}''$ explored in this paper remain consistent with the experimental lifetime measurement at this time.

As mentioned in the introduction, one of the major naturalness problems that one faces with the introduction of leptoquarks (even after the elimination of the terms that violate $B$ and $L$) is their contribution to FCNCs. Because the couplings $\lambda_{ij}$ and $\tilde\lambda_{ij}$ are arbitrary, the leptoquarks can mediate new flavor-changing processes at rates far beyond those allowed by the SM. In order to have leptoquarks anywhere near the weak scale, one must impose on the couplings some structure that, once the quark and lepton fields have been rotated to their mass eigenbasis, do not mix quarks of different generations at the tree level. 

Such a requirement would be highly unnatural unless the leptoquark interactions knew about the structure of Yukawa interactions and were somehow aligned with them. This is precisely what one finds in the Minimal Flavor Violation scheme. In MFV, one promotes the approximate $SU(3)$ flavor symmetries of the quarks (and in some cases also the leptons) to exact symmetries that are broken only by the Yukawa interactions, which one treats as spurions. In such a case, the diagonalization of the quark masses similarly diagonalizes all other flavor symmetry-breaking terms, up to corrections due to the off-diagonal elements of the CKM matrix.

In order to impose MFV on our Lagrangians, we must make several assumptions, which we now outline: 

\begin{itemize}
\item We will require the leptoquark interaction terms to be singlets under the SM flavor symmetry group in the quarks sector, $\it i.e.$, $\mathcal{G}_q \equiv SU(3)_Q\times SU(3)_u\times SU(3)_d$. Doing this imposes a symmetry structure on the leptoquarks and coupling constants. The leptoquarks themselves must transform under $\mathcal{G}_q$ and the coupling constants must be written as an expansion in the Yukawa matrices $Y_U$ and $Y_D$, which transform under $\mathcal{G}_q$ as $(\mathbf{3}, \bar{\mathbf{3}}, \mathbf{1})$ and $(\mathbf{3}, \mathbf{1}, \bar{\mathbf{3}})$, respectively. Thus, we are working with a model that exhibits a multiplicity of leptoquarks, only a few of which will be relevant for explaining the $R_{D^{(*)}}$ anomaly.

\item We will require that the leptoquarks couple only to the $\tau$ lepton, a requirement sometimes called ``$\tau$ alignment". In principle, we could try to perform an MFV-like analysis in the lepton sector, but this is both unnecessarily complicated and also not unique given our lack of understanding of the source(s) for neutrino masses. By doing this, we are also assuming that the solution to the $\rdrdstar$ anomaly is entirely due to new physics contributions to the $b\to c\tau\bar\nu_\tau$ process, with no new contributions to $b\to c\ell\bar\nu_\ell$ for $\ell = e,\mu$. 

\end{itemize}

Under these assumptions, our Lagrangians reduce to the following forms for the $S_1$ and $S_3$ leptoquarks, respectively:
\begin{align}\label{Lag}
{\cal L}_{S_1} = &\lambda_{1L} (S_1 Y)^i \bar Q^c_i\, i\tau_2\, L_3 + \lambda_{1R} (S_1 Y')^k \bar u^c_k\, \tau +h.c.\\ \nonumber
 &~~ \\ 
{\cal L}_{S_3} = &\lambda_3 (S_3^a\, Y'')^i \bar Q^c_i\, i\tau_2\tau_a\, L_3 + h.c.
\end{align}
Here the $\lambda_i$ $(i=1L,1R,3)$ are the overall couplings that multiply ${\cal G}_q$-invariant terms; we shall refer to these as the ``universal" couplings henceforth. $L_3$ is the 3rd-generation lepton doublet, and $\tau$ the right-handed charged tau lepton. The $Q$ and $u$ fields are the $SU(2)$ doublet and singlet quarks, and their indices ($i$ or $k$) are $SU(3)_Q$ and $SU(3)_u$ flavor indices, respectively, indicating that $Q$ and $u$ are taken to transform as fundamentals under their respective flavor groups. The terms set off in parenthesis, such as $(S_1 Y)^i$, are contractions of the leptoquark flavor multiplets with some number of Yukawa matrices (here collectively denoted $Y$) that serve as spurions under ${\cal G}_q$. The products $(S_1 Y)$ and $(S_3 Y'')$ transform as a $(\bar 3,1,1)$ under ${\cal G}_q$, while $(S_1 Y')$ transforms as a $(1,\bar 3,1)$. The contractions between the leptoquark field and the Yukawa spurions can be quite complicated; we do not confine ourselves to the often-used simplification that $Y$ transforms simply as an adjoint of one of the three $SU(3)$ flavor groups. This allows us to probe flavor structures that are often ignored in MFV analyses.

In general, the $Y$ (or $Y',Y''$) term that appears in the equations above is a linear combination of an arbitrary number of product of Yukawa matrices:
\beq
Y \sim a_0 + a_1 Y_u + a_2 Y_u Y_u^{\dagger} + \cdots + b_2 Y_d Y_d^{\dagger} + \cdots + c_1 Y_u Y_d + \cdots
\label{MFVex}
\eeq
where some of the coefficients $a_i, b_i, c_i$ are zero depending on how the leptoquark transforms under ${\cal G}_q$, and depending on whether contractions between the spurions, the leptoquark, and the quark doublet can generate ${\cal G}_q$-invariant terms as required in our analysis.

Having now a form for our Lagrangian, we need to identify all possible representations of ${\cal G}_q$ under which the leptoquark fields can transform. Here we impose an additional constraint:
\begin{itemize}
\item Because of the potentially large numbers of leptoquarks being introduced, all of which transform as triplets under QCD, asymptotic freedom could be lost in our theory. Therefore, we will require the total number of QCD triplets not to exceed 16, at which point the one-loop QCD $\beta$-function flips sign. This requirement will limit the number of possible representations that we need to consider, though it is not an absolute requirement for a self-consistent, low-energy theory.
\end{itemize}
Provided this, we find that  $S_1$ and $S_3$ can only have the following quantum numbers under $\mathcal{G}_q$ (see Appendix \ref{app1} for more details):
\begin{align}\label{Gq}
S_1: ~&(\bar{3},1,1), (1,\bar{3},1), (1,1,\bar{3}) \nonumber \\
		  & (6,1,1), (1,6,1), (1,1,6) \nonumber \\
		  & (3,3,1), (3,1,3), (1,3,3)\\
		  \nonumber \\
S_3: ~&(\bar{3},1,1), (1,\bar{3},1), (1,1,\bar{3}). \nonumber
\end{align}

From this point on, these quantum numbers under $\mathcal{G}_q$ will be referred to as flavor charges. In the analysis that follows, we will calculate the Wilson coefficients, $C_{S_L}''$ and $C_{S_R}''$, by integrating out leptoquarks with any of the above flavor charges. In addition to $C_{S_L}''$ and $C_{S_R}''$, we also calculate the Wilson coefficient for the operator $(s^c P_R \bar{\nu})(\bar{b}^c P_L \nu)\equiv \mathcal{O}_{bs\nu\bar\nu}$, as it generates the decay $b\to s \nu \bar{\nu}$, and will be used in later sections to check the feasibility of MFV models as an explanation of the $\rd$ anomalies.

Finally, to guarantee the self-consistency of the MFV framework \cite{Colangelo:2008qp}, we require the following:
\begin{itemize}
\item The coefficients $a_i, b_i, c_i$ in the expansion of Eq.~(\ref{MFVex}) are $O(1)$ or smaller, whereas the overall couplings $\lambda_{i}$ in Eq.~(\ref{Lag}) cannot be larger than $\sqrt{4\pi}$.
\end{itemize}
The reason for the above requirements is simple. In the expansion of Eq.~(\ref{MFVex}), if we were to allow the coefficients to become arbitrarily large, then we could generate essentially any form desired for $Y$. In that sense, MFV would provide no real constraint on the form of our couplings, and would provide no real protection against large FCNCs. Similarly, if we allowed the $\lambda_i$ to become arbitrarily large, perturbativity of the theory would at some point break down for at least some allowed couplings in the expansion. Thus we limit $|\lambda_i|^2$ to be, conservatively, less than $4\pi$. 

In order to calculate the Wilson coefficients, we begin by choosing one of the allowed flavor charges from (\ref{Gq}) and rewrite its Lagrangian using the expansion of Eq.~(\ref{MFVex}). We include up to 6 powers of each of the Yukawa matrices ({\it i.e.} $Y_d,~Y_u^{\dagger},~Y_d,~Y_d^{\dagger}$) in the expansion, as higher powers would only lead to a rescaling of the overall couplings $\lambda_{i}$. We find that only a fraction of these expansion terms can form a $\mathcal{G}_q$-singlet. For each of the terms that can form a singlet, we then find all the possible contractions of the $SU(3)$ flavor indices, using combinations of $\delta^i_j$ and $\epsilon^{ijk}$ (and/or $\epsilon_{ijk}$) for each $SU(3)$.
Among the list of possible contractions, we only consider the contractions that give the dominant contributions to the Wilson coefficients $C_{S_R}''$ (or $C_{S_L}''$). In general, the dominant contractions are the ones that require the least number of Yukawa matrices to obtain a $\mathcal{G}_q$-invariant interaction, but this need not always be the case (see, for example, the Lagrangian of $(1,3,3)$ in Table~\ref{tab:WC3_OSR}).

Table~\ref{tab:WC3_OSR} summarizes the leading terms in the Lagrangian for $S_1$ and $S_3$ for all the allowed flavor charges given in (\ref{Gq}), and associated Wilson coefficients for the operators $\mathcal{O}_{S_R}''$ and $\mathcal{O}_{bs\nu\bar\nu}$. In order to avoid repetition and make the table concise, we have omitted a factor of $\lambda_i$ (where $i=1L,3$) from the Lagrangian terms and common factors of $|\lambda_i|^2/M_{S_i}^2$ from the Wilson coefficients.
In order to be as explicit as possible with the flavor group contractions, the flavor indices are shown explicitly: the indices of $SU(3)_Q$ are denoted by $q_i$, $SU(3)_u$ by $u_i$, and $SU(3)_d$ by $d_i$. Products of Yukawa matrices in parenthesis are simple matrix multiplications, in the order shown.

A similar table can be given for $\mathcal{O}_{S_L}''$. However, one finds when making such a table that all the coefficients, $C_{S_L}''$, are highly suppressed by small Yukawa couplings; we will explain this claim in more detail in the next section. This implies that the contributions of $S_1$ to $C_{S_L}''$ are always small, and therefore have no impact on the $\rd$ anomaly nor are they constrained by the $B_c$ lifetime.

We now look at the constraints on the flavor charges for $S_1$ and $S_3$ and find those cases that can explain the $R_{D^{(\!*\!)}}$ anomaly under the assumption of MFV. 

\begin{table*}[!ht]
	\begin{ruledtabular}
		{\renewcommand{\arraystretch}{1.5}
			\begin{tabular}{c  c  c c}
				Leptoquark & Lagrangian, $\mathcal{L}$ & 
				$C_{S_R}''$&
				$C_{bs\nu\bar\nu}$\\
				\hline \hline
				
				$S_1(\bar{3},1,1)$ & $S_1^{~q_1} \left\{(V^\dagger \bar{u}^c_L )_{q_1} \tau_L - \bar{d}^c_{L q_1} \nu_L\right\}$ & 
				$V_{cb}$ & $0$ \\
				\hline
				
				$S_1(1,\bar{3},1)$ & $S_1^{~u_1} (\yud)_{u_1}^{~ q_1} \left\{(V^\dagger \bar{u}^c_L )_{q_1} \tau_L - \bar{d}^c_{L q_1} \nu_L\right\}$ & $V_{cb} y_c^2$ & $-V_{ts} V_{tb} y_t^2$\\
				\hline
				$S_1(1,1,\bar{3})$ & $S_1^{~d_1} (\ydd)_{d_1}^{~ q_1} \left\{(V^\dagger \bar{u}^c_L )_{q_1} \tau_L - \bar{d}^c_{L q_1} \nu_L\right\}$ & $V_{cb} y_b^2$ &  $0$\\
				\hline
				$S_1(3,3,1)$ & $S_{1\, q_1 u_1} (Y_u)_{q_2}^{~ u_1} \epsilon^{q_1 q_2 q_3} \left\{(V^\dagger \bar{u}^c_L )_{q_3} \tau_L - \bar{d}^c_{L q_3} \nu_L\right\} $ & $V_{cb} y_t^2$ &$V_{ts} V_{tb} y_t^2$\\
				\hline
				
				$S_1(1,3,3)$ & $S_{1\, u_1 d_1} (\yud Y_u \yud Y_d)_{u_2}^{~ d_1} (Y_u^\dagger)_{u_3}^{~ q_1}
				\epsilon^{u_1 u_2 u_3} \left\{(V^\dagger \bar{u}^c_L )_{q_1} \tau_L - \bar{d}^c_{L q_1} \nu_L\right\} $ & $V_{cb} y_c^2 y_b^2 y_t^6$ & $V_{ts} V_{tb} y_b^2 y_c^2 y_t^6$\\

				& $S_{1\, u_1 d_1} (Y_d)_{q_1}^{~d_1}(Y_u)_{q_2}^{~ u_1}\epsilon^{q_1 q_3 q_2} \left\{(V^\dagger \bar{u}^c_L )_{q_3} \tau_L - \bar{d}^c_{L q_3} \nu_L\right\}$ & $V_{cb} V_{us}^2 y_s^2 y_t^2$ & $V_{ts} V_{tb} y_d^2 y_t^2$\\
				
				\hline
				$S_1(3,1,3)$ & $S_{1\, q_1 d_1} (\yud Y_d)_{u_1}^{~d_1}(Y_u)_{q_2}^{~ u_1} \epsilon^{q_1 q_3 q_2} \left\{(V^\dagger \bar{u}^c_L )_{q_3} \tau_L - \bar{d}^c_{L q_3} \nu_L\right\} $ & $V_{cb} y_b^2 y_t^4 $ & $V_{ts}  V_{tb}^3 y_b^2 y_t^4$\\

				\hline
				$S_1(6,1,1)$ & $S_{1\, q_1 q_2} (Y_u \yud)_{q_3}^{~ q_1} \epsilon^{q_2 q_4 q_3} \left\{(V^\dagger \bar{u}^c_L )_{q_4} \tau_L - \bar{d}^c_{L q_4} \nu_L\right\} $ & $V_{cb} y_t^4 $ &  $V_{ts} V_{tb}^3  y_t^4$\\

				\hline
				$S_1(1, 6,1)$ & $S_{1\, u_1 u_2} (\yud Y_u)_{u_3}^{~ u_1}(Y_u^\dagger)_{u_4}^{~ q_1}  \epsilon^{u_2 u_4 u_3} \left\{(V^\dagger \bar{u}^c_L )_{q_1} \tau_L - \bar{d}^c_{L q_1} \nu_L\right\} $ & $V_{cb} y_c^2 y_t^4 $ &  $ V_{ts} V_{tb} y_c^2 y_t^4$\\
				
				\hline
				$S_1(1, 1,6)$ & $S_{1\, d_1 d_2} (Y_d)_{q_1}^{~d_1} (Y_u \yud Y_d)_{q_2}^{~d_2}   \epsilon^{q_1 q_3 q_2} \left\{(V^\dagger \bar{u}^c_L )_{q_3} \tau_L - \bar{d}^c_{L q_3} \nu_L\right\} $ & $-V_{td} V_{cd}V_{tb}^3 y_s^2 y_b^2 y_t^4$ & $ V_{ts} V_{tb}^3 y_d^2 y_b^2 y_t^4$\\
				\hline \hline
				
				$S_3(\bar{3},1,1)$ & $S_3^{1/3~q_1} \left\{(V^\dagger \bar{u}^c_L )_{q_1} \tau_L + \bar{d}^c_{L q_1} \nu_L\right\}$ & 
				$-V_{cb}$ &  $0$ \\
				\hline
				
				$S_3(1,\bar{3},1)$ & $S_3^{1/3~u_1} (Y_u^\dagger)_{u_1}^{~ q_1} \left\{(V^\dagger \bar{u}^c_L )_{q_1} \tau_L + \bar{d}^c_{L q_1} \nu_L\right\}$ & $-V_{cb} y_c^2$ &  $-V_{ts} V_{tb} y_t^2$\\
				\hline
				$S_3(1,1,\bar{3})$ & $S_3^{1/3~d_1} (Y_d^\dagger)_{d_1}^{~ q_1} \left\{(V^\dagger \bar{u}^c_L )_{q_1} \tau_L + \bar{d}^c_{L q_1} \nu_L\right\}$ & $-V_{cb} y_b^2$ & $0$

				\rule{0pt}{3ex} 
			\end{tabular}
		}
	\end{ruledtabular}
	\caption{\label{tab:WC3_OSR}
		Leading terms in the MFV Lagrangian for leptoquarks $S_1$ and $S_3$, along with the Wilson coefficients for the operators $\mathcal{O}_{S_R}''$ and $\mathcal{O}_{bs\nu\bar\nu}$. The first column indicate how each leptoquark transforms under the flavor group $\mathcal{G}_q\equiv SU(3)_Q\times SU(3)_u \times SU(3)_d$.}
\end{table*}

\section{\label{section4} Constraints \& Results}

We see from Table~\ref{tab:WC3_OSR} that there are 9 (3) choices of flavor charges for $S_1$ ($S_3$) that can explain the $R_{D^{(*)}}$ anomaly under the assumption of MFV. But the very presence of MFV leads to other non-trivial operators that are highly constrained. In this section, we find the constraints on these operators stemming from: (1) consistency of MFV 
(2) limits on $b\to s \nu \bar{\nu}$, and
(3) electroweak precision measurements.

\subsection{\label{section41}Consistency of MFV}

The Wilson operator analysis by itself points to leptoquarks with masses in the range of a few hundred GeV to a few TeV, with couplings that are $O(1)$. But, in some cases, imposing MFV forces the universal couplings $\lambda_{i}$ to be extremely large in order to solve the $\rdrdstar$ anomaly.

As we discussed in the previous section, we impose a rather conservative bound on the $\lambda_i$ and on the coefficients in the MFV expansion: $|\lambda_i|^2\leq 4\pi$ and all other coefficients $\leq 1$. Such a bound, surprisingly, immediately excludes the operator $\mathcal{O}_{S_L}''$ as a solution to the anomaly.
Following the procedure outlined in the last section, we find that the coefficients $C_{S_L}''$ for {\it all}\/ flavor charges of $S_1$ are suppressed by a factor of $y_c V_{cb}$ or smaller. Such coefficients would need universal couplings ($\lambda_i$) of $O(40)$ in order to explain the anomaly, which are clearly excluded by our assumptions. For example, if we take all coefficients in the MFV expansion (Eq.~(\ref{MFVex})) to equal 1, then, in order to explain the anomaly, we need $|C_{S_L}''|=|(\lambda_{1R} \lambda_{1L} y_c V_{cb})/M_{S_1}^2|\sim0.43/\tev^2$ (from Eq.~(\ref{CSLpp})). This requires $\sqrt{\lambda_{1R} \lambda_{1L}} \sim 38$ for $M_{S_1}=1 \tev$, which clearly violates our assumption that the $\lambda_i$ must remain perturbative. 

This bound also excludes the flavor charges $(1,\bar{3},1)$, $(1,3,3)$, $(1,6,1)$ and $(1,1,6)$ for $S_1$, as their contributions to $C''_{S_R}$ are suppressed by either $y_c^2$ or $y_s^2$ (see Table~\ref{tab:WC3_OSR}) at leading order. For example, consider the $(1,\bar 3, 1)$ flavor charge. In this case the $C''_{S_R}$ coefficient has the form $\lambda_{1L}^2 y_c^2 V_{cb}/M_{S_1}^2$. From Eq.~(\ref{CSRpp}), in order to explain the $\rdrdstar$ anomaly we need $C''_{S_R} = 0.293$. This means that for $M_{S_1}=1\tev$, one would need $\lambda_{1L} \simeq 80$, which is again ruled out by our requirements.

As an additional surprise, this constraint also rules out {\it all}\/ flavor charges for $S_3$, though this is not immediately apparent from Table~II. One sees in Table~II that the $S_3$ flavor charges all enforce that $C''_{S_R}<0$. In this case, according to Eq.~(\ref{CSRpp}), one is forced to a limit in which $|C''_{S_R}| \simeq 5.75/$TeV$^2$, and a suppression of $V_{cb}\simeq 0.04$ pushes $\lambda_{3}\gsim 12$, which again violates our assumptions. 

In principle, coefficients suppressed by powers of $y_b$ are also excluded (\eg, $(1,1,\bar 3)$), at least with the Higgs sector of the SM. However, it is well known that in two-Higgs doublet models
it is possible for $y_b\sim O(1)$ in the large $\tan\beta$ limit. Thus we keep operators
that scale as powers of $y_b$ to account for the possibility of an extended Higgs sector.

Thus, imposing the constraint that the universal couplings $\lambda_i$ are perturbative and the coefficients in the MFV expansion are $O(1)$ or smaller reduces our analysis to just few remaining cases: a leptoquark $S_1$ with flavor charges $(\bar 3,1,1)$, $(1,1,\bar 3)$, $(6,1,1)$, $(3,3,1)$, or $(3,1,3)$.

\subsection{\label{section42}$b\to s \nu \bar{\nu}$}

The 3-body decay $b\to s \nu \bar{\nu}$ is a flavor-changing neutral current, and thus is GIM suppressed in the SM.  Leptoquarks, however, can mediate the $b\to s \nu \bar{\nu}$ transition at tree-level, and so measurements of this process can strongly constrain the leptoquarks' parameter space. Among the exclusive processes mediated by this decay, the strongest constraint comes from the branching fraction of $B^+ \to K^+ \nu \bar{\nu}$ in the SM. The effective Hamiltonian for the $b\to s \nu \bar{\nu}$ transition in the SM can be written \cite{Buras:2014fpa}:
$$
\mathcal{H}^{\rm SM}_{\rm eff} = \sum\limits_{i=e,\mu,\tau} \, C_{\rm SM}(\bar{s}\gamma_\mu P_L b)(\bar{\nu_i}\gamma^\mu P_L \nu_i) \nonumber
$$ 
where 
$$ C_{\rm SM} = 4 \sqrt{2} G_F V_{tb}V_{ts}^*\frac{e^2}{16\pi^2}\frac{X_t}{s_w^2},~~X_t=1.469 \pm 0.017.$$
This leads to a predicted branching ratio of \cite{Buras:2014fpa}:
\beq
 \mathcal{B}(B^+ \to K^+ \nu \bar{\nu})_{\rm SM}=(3.98 \pm 0.43 \pm 0.19) \times 10^{-6}
 \eeq
  while experiment places an upper bound on this process at \cite{Lees:2013kla}:
  \beq
  \mathcal{B}(B^+ \to K^+ \nu \bar{\nu})_{\rm Exp}<1.6 \times 10^{-5}.
  \eeq

The corresponding effective Hamiltonian obtained by integrating out $S_1$ can be written:
$$
\mathcal{H}^{S_1}_{\rm eff}= C_{bs\nu\bar\nu}(\bar{\nu}_\tau P_R s^c)(\bar{b}^c P_L \nu_\tau) 
$$
which, upon Fierzing, gives
$$
\mathcal{H}^{S_1}_{\rm eff}= \frac{C_{bs\nu\bar\nu}}{2} (\bar{s}\gamma_\mu P_L b)(\bar{\nu}_\tau\gamma^\mu P_L \nu_{\tau}).
$$
This operator can interfere with the SM contribution for $\nu_i=\nu_\tau$, modifying the $b\to s \nu \bar{\nu}$ decay rate. In order to obtain bounds on $C_{bs\nu\bar\nu}$, we calculate the ratio,
\beq
R = \frac{\mathcal{B}(B^+ \to K^+ \nu \bar{\nu})_{\rm SM}}{3\, C_{\rm SM}^2}
\eeq 
and estimate the new branching ratio to be
\beq
\mathcal{B}(B^+ \to K^+ \nu \bar{\nu})= R\left(\left(C_{\rm SM}+\frac{C_{bs\nu\bar\nu}}{2}\right)^2+ 2 \,C_{\rm SM}^2\right).
\eeq
For this branching ratio to remain below the experimental bound, we find that
\beq
-0.045<C_{bs\nu\bar\nu} (\text{TeV}^{-2}) < 0.087.
\eeq

For $S_1$ flavor charges $(3,3,1)$, $(3,1,3)$, and $(6,1,1)$, one finds that $C_{bs\nu\bar\nu} \simeq -C_{S_R}''$ (see Table~\ref{tab:WC3_OSR}), which means that in order to explain the $\rd$ anomaly, $C_{bs\nu\bar\nu} \simeq -0.315/\tev^2$. Since this value is well outside the experimentally allowed range, we conclude that the $(3,3,1)$, $(3,1,3)$ and $(6,1,1)$ flavor charges for $S_1$ are disallowed as solutions to the anomaly by $b\to s \nu \bar{\nu}$. It should be noted that one cannot avoid these constraints simply by decoupling some of the flavor components of the leptoquarks, because the operators $\mathcal{O}_{S_R}''$ and ${\cal O}_{bs\nu\bar\nu}$ decay are mediated by the same leptoquark flavor components.

We note that the contributions from $S_1$ to ${\cal O}_{bs\nu\bar\nu}$ are entirely due to $\lambda_{1L}$; the coupling $\lambda_{1R}$ does not contribute at all to this process. Therefore turning on $\lambda_{1R}$ and generating a non-zero ${\cal O}_{S_L}''$ can in no way help weaken this constraint. 

At this point, we are left with only two remaining options for the leptoquark flavor charges that can explain the anomaly and are not disallowed by the constraints we have studied so far: $S_1(\bar{3},1,1)$ and $S_1(1,1,\bar{3})$. 

\subsection{\label{section43}Precision electroweak observables}

The dominant effect of the $S_1$ leptoquark on precision electroweak observables is through its modification of the couplings of $Z$ to fermions at one loop; the relevant diagrams are shown in Appendix~\ref{app2}. A number of electroweak observables can be impacted by the presence of a (predominantly) third-generation leptoquark, including the invisible width of the $Z$, the forward-backward asymmetry of $Z\to\bar bb$ or $\bar\tau\tau$, or the total rates for these same two processes.
We find that the partial decay width of $Z \to \tau \bar{\tau}$ imposes the strongest constraints on $S_1 (\bar{3},1,1)$ and $S_1(1,1,\bar{3})$. Additional details on the calculation are found in Appendix~\ref{app2}. We consider each of these two flavor charges in turn below.

~

\noindent {\bf  $\mathbf{S_1 (\bar{3},1,1)}$:} Reading from Table~\ref{tab:WC3_OSR}, the relevant part of the Lagrangian for the triplet of $S_1 (\bar 3,1,1)$ leptoquarks is:
\begin{align*}
\mathcal{L} &\supset \lambda_{1L} S_1^{~q_1} \left\{(V^\dagger \bar{u}^c_L )_{q_1} \tau_L - \bar{d}^c_{L q_1} \nu_L\right\}\\
&= \lambda_{1L} S_1^{~1}\left\{(V_{ud} \bar{u}^c_L  + V_{cd}  \bar{c}^c_L + V_{td} \bar{t}^c_L)\tau_L-\bar{d}^c_{L}\nu_L\right\}\\
&~+\lambda_{1L} S_1^{~2} \left\{(V_{us} \bar{u}^c_L+ V_{cs} \bar{c}^c_L+ V_{ts} \bar{t}^c_L)\tau_L-\bar{s}^c_{L}\nu_L\right\}\\
&~+\lambda_{1L} S_1^{~3} \left\{(V_{ub} \bar{u}^c_L + V_{cb}\bar{c}^c_L + V_{tb} \bar{t}^c_L)\tau_L -\bar{b}^c_{L}\nu_L\right\}
\end{align*}
Note that among the flavor triplet of leptoquarks, only $S_1^{~3}$ contributes to $\rd$, and can explain the anomaly for $\lambda_{1L}\simeq 2.8$ (for $M_{S_1}$ = 1 TeV). We need such a large value of $\lambda_{1L}$ to make up for the $V_{cb}$ suppression in the $S_1^3 \bar{c}^c_L \tau_L $ coupling. But this implies that the $S_1^3 \bar t^c_L \tau_L$ coupling, which is only ``suppressed" by $V_{tb}$, becomes fairly large. As we will see, the large coupling constant in this particular interaction (with a $t$-quark) makes this leptoquark very sensitive to electroweak observables. 

Before calculating the constraints from electroweak precision measurements, we note that the $\tau$ lepton has a large coupling to the $u$-quark through two of the components of the leptoquark flavor triplet: $S_1^{~1}$ and $S_1^{~2}$. These components can therefore be strongly constrained by $\bar\tau\tau$ production at the LHC. However, one can avoid such constraints, if needed, by assuming $S_1^{~1}$ and $S_1^{~2}$ to be much heavier than $S_1^{~3}$, which is allowed within the framework of MFV, since the mass of $S_1^{~3}$ can split from its flavor partners due to the large top Yukawa \cite{Grossman:2007bd}. 
For completeness, we will calculate below the electroweak constraints both with and without decoupling the leptoquark components $S_1^{~1}$ and $S_1^{~2}$.

If we assume for now that the flavor triplet of leptoquarks is degenerate, then the shift in the coupling of the $Z$ to leptons can be expressed as: 
\begin{align}\label{deltagS1Ztau}
&\Delta g_{L}^{S_1} (Z\rightarrow\bar\tau\tau)  =  \nonumber\\
&~~~\frac{3 g_2|\lambda_{1L}|^2 m_t^2}{32 \pi^2 c_w M^2} \left(2 \log \frac{M}{m_t} - 1 \right)
+2\frac{ g_2|\lambda_{1L}| ^2m_Z^2}{96\pi ^2 c_w M^2} \times\nonumber\\
& \left\{ \left(-\frac{1}{2}+s_w^2\right)-\left(\frac{1}{2}-\frac{2}{3} s_w^2\right) \left(12 \log\frac{M}{m_Z}+1+ \boldsymbol{i}6\pi\right)\right\}
\end{align}
\begin{align}\label{deltagS1Znu}
&\Delta g_L^{S_1} (Z\rightarrow \bar\nu\nu) = 3\frac{g_2|\lambda_{1L}| ^2m_Z^2}{96 \pi ^2 c_w M^2}\times\nonumber\\
&~~~ \left\{ \frac{1}{2}-\left(-\frac{1}{2}+\frac{1}{3} s_w^2\right) \left(12 \log\frac{M}{m_Z}+1+ \boldsymbol{i}6\pi\right)\right\}
\end{align}
For the $S_1(\bar 3,1 ,1)$ leptoquark, there is no equivalent correction to the right-handed couplings of the $Z$ to fermions due to the structure of our Lagrangian.
These relations are valid up to leading order in $m_Z/M$ and $m_t/M$ (where $M$ is the mass of $S_1$).  In the expression for the correction to the $Z\bar\tau\tau$ coupling, the first term dominates and is due to the contribution of the top quark in the loop, where it picks up an enhancement $\propto m_t^2$ from helicity flips on each of the $t$-quark lines. The second, smaller contribution is due to $u$- and $c$-quarks running in the loop, and includes an imaginary component when the quarks go on shell. This smaller contribution has the opposite sign to the dominant term and is thus included in our calculations in order to obtain conservative bounds. At the same time, the shift in the $Z$ coupling to $\bar\nu\nu$ is due to down-type quarks in the loop, and picks up no large enhancements. (Shifts in the $Z$ couplings to quarks involve only leptons in the loops and are even smaller.) Note that in Eq.~(\ref{deltagS1Ztau}), we have omitted a term proportional to $m_Z^2/M^2$ (which appears in Eq.~(\ref{AppdeltagmL})) because of its negligible effect.

\begin{figure}[t]
		\centering
		\includegraphics[width=.95\linewidth]{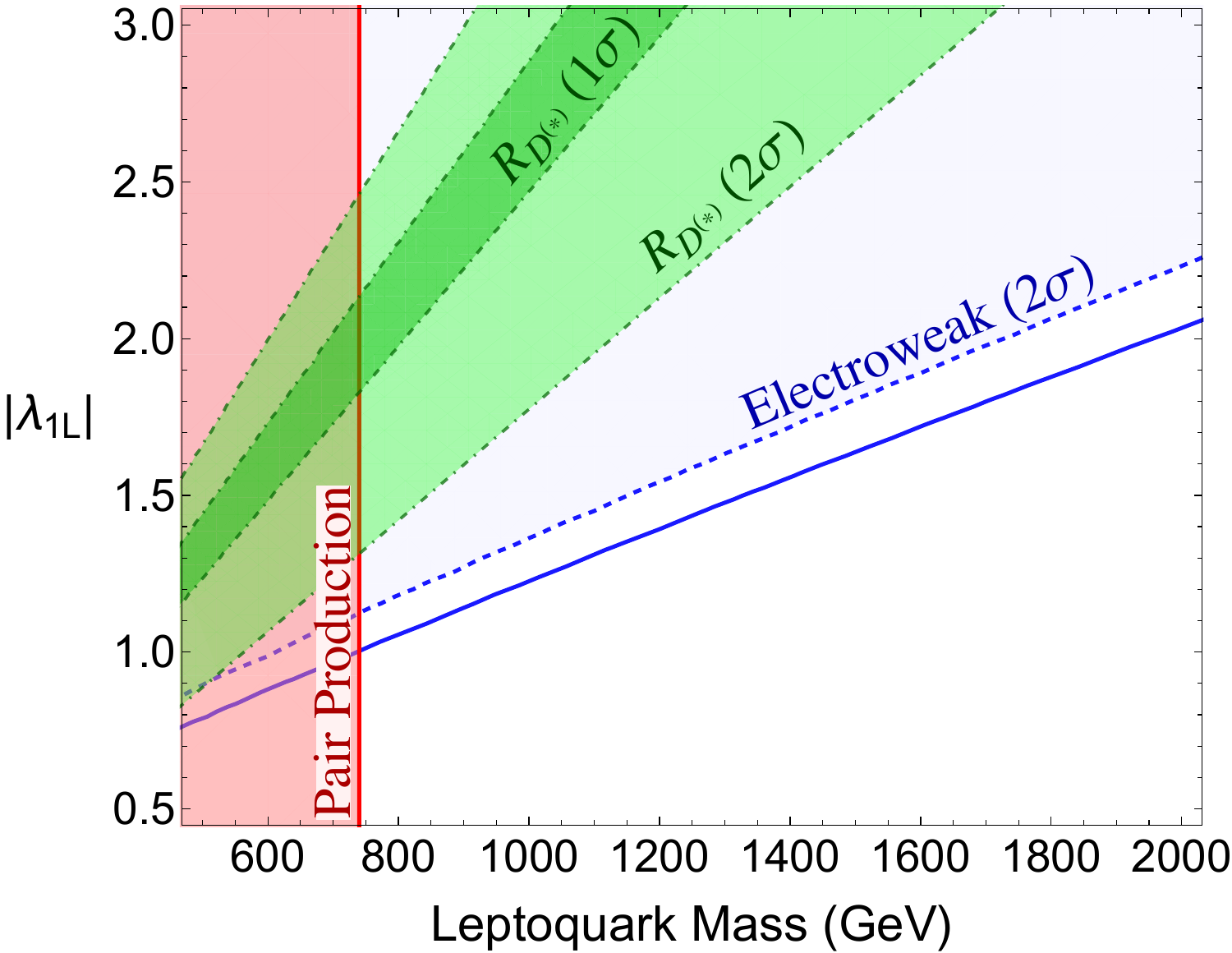}
	\caption{\label{fig:S1311_OSR_EW} Parameter space of the leptoquark $S_1(\bar{3},1,1)$ in which it can explain the $\rdrdstar$ anomaly (green shaded region), along with the regions excluded by electroweak precision measurements (above dashed blue line) and pair production bounds (left of solid red line). The solid blue line indicates the electroweak constraints after decoupling the $S_1^{1,2}$.  }
\end{figure}

Figure~\ref{fig:S1311_OSR_EW} summarizes the electroweak constraints on $S_1(\bar{3},1,1)$. The green region is the parameter space where the $\rd$ anomaly can be explained by this choice of flavor charge, within either $1\sigma$ (or $2 \sigma$) of the experimental measurements. Meanwhile, the constraints from the electroweak data disfavor the region above the dashed blue line in Fig.~\ref{fig:S1311_OSR_EW} at 95\% C.L. (\ie, $\Delta\chi^2\geq5.99$). These constraints are obtained by a $\chi^2$ fit to all the electroweak observables that are strongly affected by the presence of leptoquarks. These observables include $R_\tau$, $\Gamma(\text{inv})$, $R_b$, $A_\tau$ and $A_b$ (see Appendix~\ref{app2} for the definitions of these observables). The current direct pair production bounds from LHC are also indicated by the red vertical line.

If we assume that the leptoquarks $S_1^1$ and $S_1^2$ are much heavier than $S_1^3$, then the electroweak constraints become even stronger, with the new 95\% C.L. now indicated by a solid blue line in Fig.~\ref{fig:S1311_OSR_EW}. This is because the contributions from $u$- and $c$-quarks to $\Delta g_{L}^{S_1} (Z\rightarrow\bar\tau\tau)$ have the opposite sign to that from the $t$-quark and, after decoupling $S_1^1$ and $S_1^2$, those negative contributions become suppressed.

As Fig.~\ref{fig:S1311_OSR_EW} demonstrates, there is no parameter space remaining in which $S_1(\bar{3},1,1)$ can explain the $\rd$ anomaly while evading electroweak constraints. This is true either in the case with decoupled $S_1^{~1}$ and $S_1^{~2}$, or without.

~

\noindent {\bf $\mathbf{S_1 (1,1,\bar{3})}$:} The contributions from these leptoquarks to $\mathcal{O}_{S_R}''$ mimic exactly those of the previous case, but now with $\lambda_{1L}$ replaced by $\lambda_{1L}\, y_b$. As such, they are highly suppressed for SM-like bottom Yukawa couplings, but  could be sizable in a two-Higgs doublet model with large $\tan\beta$, where the bottom Yukawa can be $O(1)$. In either case, the leading term in the Lagrangian of $S_1(1,1,\bar{3})$, taken from Table~\ref{tab:WC3_OSR}, can be written as:
\begin{align*}
\mathcal{L} &\supset \lambda_{1L} y_b S_1^{~3} \left\{(V_{ub} \bar{u}^c_L + V_{cb}\bar{c}^c_L + V_{tb} \bar{t}^c_L)\tau_L -\bar{b}^c_{L}\nu_L\right\},
\end{align*}
ignoring terms suppressed by $y_d$ and $y_s$. 
For $y_b=1$, this Lagrangian is exactly that of the $S_1^{~3}$ component of $S_1(\bar{3},1,1)$. This implies that Fig.~\ref{fig:S1311_OSR_EW} can also be used to study the parameter space of $S_1(1,1,\bar{3})$, with $|\lambda_{1L}|$ on the y-axis replaced by $|\lambda_{1L}y_b|$ and constraints from electroweak precision measurements denoted by the solid blue line. As before, there is no parameter space for $S_1 (1,1,\bar{3})$ where it can both avoid the electroweak constraints and explain the $\rd$ anomaly.

~

\noindent {\bf Other flavor charges:} In this section, we have concentrated our discussion on the two leptoquarks that survived the constraints on the consistency of the MFV expansion and from $b\to s\bar\nu\nu$. However, it is worth taking a moment to indicate the effects of precision electroweak constraints on the leptoquarks previously excluded. 

Among the $S_1$ leptoquarks, the cases in which the $S_1$ transforms as a $\bar 3$ under one (and only one) of the flavor groups all behave similarly. In all cases, the coupling of the $S_1$ to $\bar c^c_L \tau_L$ is suppressed by $V_{cb}/V_{tb}$ compared to the coupling of the same leptoquark state to $\bar t^c_L \tau_L$. As such, one will always reproduce bounds similar in shape (though with rescaled $y$-axis) as those shown in Fig.~\ref{fig:S1311_OSR_EW}. Thus, even if we ignored our previous constraints, all three of these cases would be excluded by the precision electroweak constraints.

However, if the $S_1$ leptoquark is found in a $(\bar 3, 3)$ flavor representation under two of the flavor groups, or if it is in a $6$ under one of the flavor groups, then those same suppressions of the $\bar c^c_L \tau_L$ coupling relative to $\bar t^c_L \tau_L$ are not present. In these cases, the effect of the leptoquark on the $Z$ couplings is always quite small, even when the $\rd$ anomaly is explained. Such cases are therefore not ruled out by the precision electroweak data. 

In addition, we find that the precision electroweak data rules out all three possible flavor charges for $S_3$. Like the case of the $S_1(\bar 3,1,1)$, the couplings to the leptoquark to top quarks is enhanced relative to the coupling to charm. In fact, the electroweak constraints on this case are actually somewhat stronger than for the $S_1$ case, mostly due to a strong bound coming from the invisible width of the $Z$ (specifically $Z\to\bar\nu_\tau\nu_\tau$), not present in the $S_1$ case. This is described in more detail in Appendix B.

Finally, we note that turning on the coupling $\lambda_{1R}$ for the $S_1$ leptoquark does not help weaken these constraints, but rather strengthens them overall. It is true that turning on $\lambda_{1R}$ {\it does}\/ weaken the constraint from $A_\tau$, since having both left-handed and right-handed operators reduces the amount of parity violation present, of which $A_\tau$ is a measure. However, the constraint from $R_\tau$ is significantly strengthened, since both operators contribute to the rate for $Z\to\bar\tau\tau$ but cannot interfere with each other. This is discussed in more detail in Appendix B.

~

In short, we find that electroweak precision measurements provide a strong constraint on the parameter space available to solve the $\rd$ anomaly. That constraint is strong enough to rule out the two remaining cases not already ruled out by other considerations. This leaves us with no scalar leptoquarks capable of solving the $\rd$ anomaly within a completely self-consistent MFV scenario.

\section{Summary and Conclusion}

In this work, we studied the feasibility of scalar leptoquarks as an explanation of the $\rd$ anomaly under the assumption of Minimal Flavor Violation (MFV). We found that there are two scalar leptoquarks, $S_1$ and $S_3$, that generate Wilson coefficients that improve the fit of the theoretical predictions of $\rd$ with the data.

We considered all possible quantum numbers under the flavor group $\mathcal{G}_q=SU(3)_Q\times SU(3)_u \times SU(3)_d$ for these leptoquarks consistent with asymptotic freedom of QCD. This left us with a handful of models that we then analyzed. Table~\ref{tab:CominedOSR} summarizes these minimal flavor violating leptoquark models. We then proceeded to calculate the constraints on all these models from various sources. First, we required that the MFV expansion be self-consistent and perturbative, bounding the allowed couplings. Several models were thereby ruled out, and some others remain viable only within the context of a two-Higgs doublet model at large $\tan\beta$.

Along with the requirement that the MFV expansion be self-consistent, we looked at the constraints on the MFV leptoquark models from $b\to s \nu\bar\nu$ and electroweak precision measurements. We found that some models were ruled out by either of these constraints, or both. This means that all possible flavor quantum numbers for the scalar leptoquarks are ruled out as an explanation of the $\rd$ anomaly within this scenario. 

These results are summarized in Table~\ref{tab:CominedOSR}. Under our assumptions, none of the flavor charges of scalar leptoquarks can avoid the experimental constraints while explaining the $\rdrdstar$ anomaly. If we relax our assumption on perturbativity and consistency of the MFV expansion, then $S_1(1,3,3)$ and $S_1(1,1,6)$ can potentially explain the anomaly for couplings of $O(100)$ for $M_{S_1}=1 \tev$ and large $\tan\beta$. For the other flavor charges, unless there is some form of cancellation among different contractions, one cannot simultaneously avoid the experimental constraints and explain the anomaly. We conclude that, if scalar leptoquarks are solution to the $\rdrdstar$ anomaly, then in addition to violation of lepton flavor universality, this anomaly is also hinting towards a new flavor structure in the quark sector that must somehow survive well-known constraints from flavor-changing neutral currents.

\begin{table}[t]
	\begin{ruledtabular}
		{\renewcommand{\arraystretch}{1.3}
			\begin{tabular}{c  c  c c}
				Leptoquark &
				MFV consistency &
				$b\to s\bar \nu \nu$&
				Electroweak\\
				\hline \hline
				
				$S_1(\bar{3},1,1)$ & \cmark & \cmark & \xmark\\ 
				
				$S_1(1,\bar{3},1)$ & \xmark & \xmark & \xmark\\ 
				
				$S_1(1,1,\bar{3})$ & \cmark (2HDM) & \cmark & \xmark\\ 
				
				$S_1(3,3,1)$ & \cmark & \xmark & \cmark\\ 
				
				$S_1(1,3,3)$ & \xmark & \cmark & \cmark\\ 
				
				$S_1(3,1,3)$ & \cmark (2HDM) & \xmark & \cmark\\ 
				
				$S_1(6,1,1)$ & \cmark & \xmark & \cmark\\ 
				
				$S_1(1, 6,1)$ & \xmark & \xmark & \cmark\\
				
				$S_1(1, 1,6)$ & \xmark & \cmark & \cmark\\ 
				\hline\hline
				
				$S_3(\bar{3},1,1)$ & \xmark & \cmark & \xmark\\ 
				
				$S_3(1,\bar{3},1)$ & \xmark & \xmark & \xmark\\ 
				
				$S_3(1,1,\bar{3})$ & \xmark  & \cmark & \xmark\\ 
				
			\end{tabular}
		}
	\end{ruledtabular}
	\caption{\label{tab:CominedOSR}
		Summary of constraints from self-consistency of MFV, $b\to s \bar \nu \nu$, and electroweak precision measurements. A (\cmark) sign indicates that the particular flavor charge is allowed by the corresponding measurement, whereas an (\xmark) sign indicates it is not.}
\end{table}

\section*{Acknowledgments}

We thank K.S.~Babu, A.~Delgado, A.~Martin and N.~Raj 
for discussions.
This work was partially supported by the National
Science Foundation under grants PHY-1520966 and PHY-1820860.
\\

\appendix
\section{\label{app1}Possible Flavor Charges}

In this section, we find all the possible quantum numbers of leptoquarks under the flavor symmetry of quarks, $\mathcal{G}_q = SU(3)_Q\times SU(3)_u\times SU(3)_d$, that can generate a $\mathcal{G}_q$-invariant interaction for the $S_1$ and $S_3$ leptoquarks, assuming MFV. We make use of the triality of each of the $SU(3)$ groups in order to quickly classify the allowed representations. 

Let $(s_1, s_2, s_3)$ be the representations of leptoquark $S$ under $(SU(3)_Q, SU(3)_u, SU(3)_d)$. We define a list $n_{X} = (n_{X_1}, n_{X_2}, n_{X_3})$ which, for an object $X$, represents the number of fundamental indices on a given representation minus the number of anti-fundamental indices for each of the flavor $SU(3)$. This quantity, when taken (mod~3), is just the usual triality of $SU(3)$. For example, if $S$ transforms with $(s_1, s_2, s_3)=(3, \bar{3}, 8)$, then 
	$n_S = (n_{s_1}, n_{s_2}, n_{s_3})=(1, -1, 0)$.  (A value of $n_{s_i}=-1$ corresponds to the usual triality value of 2.) 

To obtain a $\mathcal{G}_q$-invariant interaction, we need to promote some combination of SM Yukawa matrices to
spurions. We define a general combination of spurions by,
\begin{align*}
	Y \equiv (Y_U)^{p_1} (Y_U^\dagger)^{p_2} (Y_D)^{p_3} (Y_D^\dagger)^{p_4} 
\end{align*}
where the $p_i$ indicate the number of copies of each spurion inserted.
Note that we have not specified the contractions of various $SU(3)$ indices of these spurions; that is, the product above is not merely matrix multiplication. In fact, this analysis is independent of the contractions among spurions and/or other fields, and the results of this section hold for all the possible contractions. For $Y_U \sim (3, \bar{3}, 1)$ and $Y_D \sim (3, 1, \bar{3})$, we get
\begin{align*}
	n_Y = (p_1-p_2+p_3-p_4, -p_1+p_2, -p_3+p_4).\nonumber
\end{align*}

With the above definitions, let us now consider an operator representing the interaction of a leptoquark with left-handed leptons and quarks, invariant under $\mathcal{G}_q$:
$$ \mathcal{O} = Y \bar{Q}^c_L S  L_L. $$
The overall triality charges for this operator are 
$$n_{\mathcal{O}}=(s_1+1+p_1-p_2+p_3-p_4, 
	s_2-p_1+p_2, s_3-p_3+p_4).$$
For this interaction to be invariant under the flavor group, $n_{\mathcal{O}}\,$(mod~3) should be equal to $0$ under each $SU(3)$. Defining, 
\begin{align*}
	s_2-u_1+u_2 \equiv 3 z_u \text{  and } s_3-d_1+d_2 \equiv3 z_d,
\end{align*}
then
\begin{align*}
	n_{\mathcal{O}}=(s_1+s_2+s_3+1 -3(z_u+z_d), 3 z_u, 3 z_d).
\end{align*}
We can enforce that the operator be invariant under $SU(3)_u$ and $SU(3)_d$ by requiring that $z_{u,d}$ be integers. For it to also be a singlet under $SU(3)_Q$ requires that $\left(s_1+s_2+s_3+1 -3(z_u+z_d)\right)\text{(mod 3)}$ should be $0$. This implies
$$	s_1+s_2+s_3 = 3z-1$$
for integer $z$. 

\begin{table}[t]
	\begin{ruledtabular}
		{\renewcommand{\arraystretch}{1.5}
			\begin{tabular}{c c}
				$\left(\sum s_i \right)\text{(mod 3)}$  & ${\cal G}_q=(SU(3)_Q, SU(3)_u, SU(3)_d)$  \\
				\hline 
				-1 & $(\bar{3},1,1)$, $(6,1,1)$, $(3,3,1)$, $(\bar{6},3,1)$, $(8,\bar{3},1)$\\
				0 & $(1,1,1)$, $(8,1,1)$, $(3,\bar{3},1)$, $(10,1,1)$, $(\overline{10},1,1)$\\
				1 & $(3,1,1)$, $(\bar{6},1,1)$, $(\bar{3},\bar{3},1)$, $(6,\bar{3},1)$, $(8,3,1)$
				\rule{0pt}{3ex} 
			\end{tabular}
		}
	\end{ruledtabular}
	\caption{\label{tab:append1}
		First few possiblities of ${\cal G}_q$-representations, $(s_1,s_2,s_3)$, for the three possible values of ($s_1+s_2+s_3$) mod 3. Note that all permutations among the flavor charge in the table are also allowed. For example, $(\bar 3,1,1)$ also implies $(1,\bar{3},1)$ and $(1,1,\bar{3})$.}
\end{table}

The lowest lying set of allowed flavor charges for $S$ for the three values of $(s_1+s_2+s_3)\text{(mod 3)}$ are given in Table~\ref{tab:append1}. Note that the permutations within each flavor charge mentioned in the table are also allowed. For example, the charge assignment $(\bar{3},1,1)$ yields $(s_1+s_2+s_3)\text{(mod 3)} = -1$, as do the permutations $(1,\bar{3},1)$ and $(1,1,\bar{3})$.

This same procedure can be used to find all allowed flavor charges for any choice of leptoquark, given the couplings present in the Lagrangian.

\section{\label{app2}Electroweak Constraints}

We now provide a more detailed discussion on using electroweak precision measurements to obtain constraints on scalar leptoquarks. This section is independent of the MFV analysis carried out in the main text. First, we will calculate the change in $Z\bar f f$ couplings due to the presence of a leptoquark coupling that violates fermion number by 2 ($\it{i.e.}$,~$\Delta F \equiv 3 \Delta B+\Delta L=2$) and then compute the electroweak contraints on $S_1$ and $S_3$ leptoquarks, assuming that they only couple to the third generation quarks and leptons. We are interested in the $F=2$ couplings because the couplings of both $S_1$ and $S_3$ leptoquarks viotate fermion number by 2.

\begin{figure}[!]	
	\centering
	\begin{subfigure}[b]{0.23\textwidth}
		\includegraphics{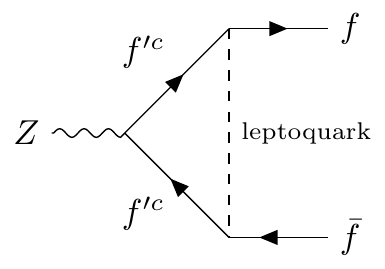} 
		\caption{}\label{FeynDiag1}
	\end{subfigure}
	\begin{subfigure}[b]{0.23\textwidth}
		\includegraphics{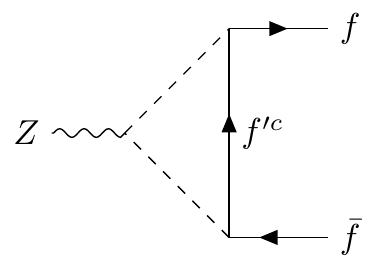} 
		\caption{}\label{FeynDiag2}
	\end{subfigure}
	\begin{subfigure}[b]{0.23\textwidth}
		\includegraphics{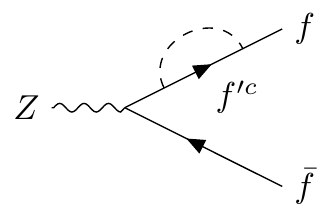} 
		\caption{}\label{FeynDiag3}
	\end{subfigure}
	\begin{subfigure}[b]{0.23\textwidth}
		\includegraphics{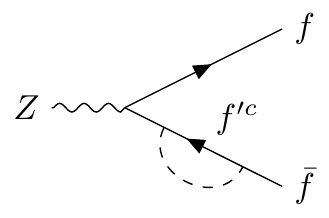} 
		\caption{}\label{FeynDiag4}
	\end{subfigure}
	\caption{\label{fig:FeynDiag}Feynman diagrams for Z decay with leptoquark.}
	\label{fig:Feyn}
\end{figure}

The interaction terms for a leptoquark, $S$, with $\Delta F=2$ can be written as:
\begin{align}
\mathcal{L}_{\Delta F=2} = \lambda_L \overline{(f'_L)^c} S f_L+\lambda_R \overline{(f'_R)^c} S f_R.
\end{align}
This leptoquark can modify the $Z\bar f f$ couplings through the Feynman diagrams shown in Fig.~\ref{fig:FeynDiag}, where $f$ is the final state fermion and $f'$ is the fermion running in the loop along with the leptoquark. Since there is a discontinuity in the amplitude of the first diagram (Fig.~\ref{FeynDiag1}) at $m_{f'} = m_Z/2$, the contributions from these diagrams can be divided into two categories: $m_{f'}\sim 0$ and $m_{f'}>m_Z/2$. All the fermions of the SM fall into the first category, except the $t$-quark, which belongs to the second case. We assume the mass of the leptoquark, $M$, to be much greater than $m_Z$ and thus there is no discontinuity in the second diagram (Fig.~\ref{FeynDiag2}). On computing the amplitudes of these diagrams, we find the dominant parts of the corrections to $Z\bar f f$ couplings for the two cases to be:

~
~

\noindent {\bf Case I:} $m_{f'}\sim 0$:
\begin{align}\label{Appdeltagmz}
&\Delta g^f_{L/R}  =\frac{g_2|\lambda_{L/R}|^2 m_Z^2}{96 \pi^2 c_w M^2 N_c^f} \times \nonumber \\
&~~~~~~~ \left\{g^f_{L/R}-g^{f'}_{L/R}\left(12 \log \frac{M}{m_Z}+1+\ii 6 \pi\right)\right\}
\end{align}

~

\noindent {\bf Case II:} $m_{f'}> m_Z/2$:
\begin{align}\label{AppdeltagmL}
&\Delta g^f_{L/R}  = \pm \frac{3 g_2|\lambda_{L/R}|^2 m_{f'}^2 T_3^{f'}}{16 \pi^2 c_w M^2 N_c^f} \left(2 \log \frac{M}{m_{f'}}-1\right) +\nonumber \\
&\frac{g_2|\lambda_{L/R}|^2 m_Z^2}{96 \pi^2 c_w M^2 N_c^f} \left\{g^f_{L/R}-g^{f'}_{L/R}\left(12 \log \frac{M}{m_{f'}}-9\right)+3 T_3^{f'}\right\},
\end{align}
where $N_c^f$ is the number of colors of the final state fermion ($f$), $g_L^{f} = T_3^{f}-Q^{f} s_w^2$ and $g_R^{f}=-Q^{f} s_w^2$. Here, $T_3^{f}$ is the the SM weak SU(2) quantum number of left handed fermion, $f$. The relations in Eqs.~(\ref{Appdeltagmz}) and (\ref{AppdeltagmL}) are valid up to leading order in $m_Z/M$ and $m_{f'}/M$. Similar calculations can be found in \cite{Mizukoshi:1994zy,ColuccioLeskow:2016dox}.

Next, we calculate the constraints from electroweak precision measurements by considering a specific structure for the leptoquarks' couplings. We will demonstrate a situation where the $S_1$ and $S_3$ leptoquarks only couple to the third generation quarks and leptons. This example will cover both the above mentioned cases for various final state fermions. In this scenario, the Lagrangian for these leptoquarks can be written:
\begin{align}\label{AppLagS1}
{\cal L}_{S_1} = &\lambda_{1L} S_1 \bar Q^c_3\, i\tau_2\, L_3 + \lambda_{1R} S_1  \bar u_3^c\, \tau +h.c.\nonumber\\
=& \lambda_{1L} S_1 (\bar t_L^c \tau_L-\bar b_L^c \nu_\tau) + \lambda_{1R} S_1  \bar t_R^c\, \tau_R +h.c.,
\end{align}
\begin{align}\label{AppLagS3}
{\cal L}_{S_3} = &\lambda_3 S_3^a \bar Q^c_3\, i\tau_2\tau_a\, L_3 + h.c.\nonumber\\
=&\sqrt{2}\lambda_3S_3^{-2/3} \bar t^c_L \nu_\tau -\lambda_3S_3^{1/3} \bar t^c_L \tau_L \nonumber\\
&-\lambda_3S_3^{1/3} \bar b^c_L \nu_\tau -\sqrt{2}\lambda_3S_3^{4/3} \bar b^c_L \tau_L+h.c.,
\end{align}
where $\tau_a (a=1,2,3)$ are the Pauli matrices, and the upper indices of $S_3$ indicate the electric charge of the corresponding leptoquark. The second equality in Eqs.~(\ref{AppLagS1}) and (\ref{AppLagS3}) is the Lagrangian after $SU(2)$ symmetry breaking, written in the mass basis of the down quark; we have ignored up-type quarks terms that are suppressed by off-diagonal CKM elements.

\begin{table}[htbp]
	\begin{ruledtabular}
		{\renewcommand{\arraystretch}{1.5}
			\begin{tabular}{cccc}
				\textrm{Observable}&
				\textrm{Experimental}&
				\textrm{Standard~Model}&
				\textrm{Pull}\\
				\hline 
				\rule{0pt}{3ex} 
				$\Gamma_Z~[GeV]$ &  2.4952 $\pm$ 0.0023 & 2.4943 $\pm$ 0.0008 & 0.4\\
				$\Gamma(had)~[GeV]$ &   1.7444 $\pm$ 0.0020 & 1.7420 $\pm$ 0.0008 & $-$\\
				$\Gamma(inv)~[MeV]$ &    499.0 $\pm$ 1.5 & 501.66 $\pm$ 0.05 & $-$\\
				$R_\tau$ & 20.764 $\pm$ 0.045 &20.779 $\pm$ 0.010 & -0.3\\
				$R_b$ & 0.21629 $\pm$ 0.00066&0.21579 $\pm$ 0.00003 &0.8\\
				$A^{(0,\tau)}_{FB}$&0.0188 $\pm$ 0.0017&0.01622 $\pm$ 0.00009& 1.5\\
				$A^{(0,b)}_{FB}$&0.0992 $\pm$ 0.0016& 0.1031 $\pm$ 0.0003& -2.4\\
				$A_\tau$&0.1439 $\pm$ 0.0043 
				& 0.1470 $\pm$ 0.0004 &-0.7\\
				$A_b$&0.923 $\pm$ 0.020 & 0.9347& -0.6\\
				
			\end{tabular}
		}
	\end{ruledtabular}
	\caption{\label{tab:AppendixEWTable}%
		The relevant LEP and SLC observables with their SM predictions \cite{Tanabashi:2018oca}. The value of $A_\tau$ corresponds to measurements at LEP using $\tau$-lepton polarization.
	}
\end{table}

The leptoquarks in this set-up predominantly modify the $Z$ decay to $\tau$,  $\nu_\tau$, and $b$. The affected LEP and SLC $Z$-pole observables with their SM predictions and measured values are summarized in Table~\ref{tab:AppendixEWTable}. In Table~\ref{tab:AppendixEWTable}, $\Gamma_Z$ is the total decay width of $Z$, $\Gamma(inv)$ is its invisible decay width, $R_\tau\equiv \Gamma(had)/\Gamma(\tau\bar{\tau})$ and $R_b\equiv \Gamma(b \bar{b})/\Gamma(had)$. $\Gamma(had)$ is the partial width of $Z$ into hadrons, which receives a new contribution from the leptoquark-mediated $Z\rightarrow b\bar{b}$ process. The terms $A^{(0,f)}_{FB}$ and $A_f$ ($f=\tau, b$) are the asymmetry observables that quantify parity violation in weak neutral currents. These are defined as:

\begin{equation}
A_f\equiv\frac{2 g_A^f g_V^f}{(g_A^f)^2 + (g_V^f)^2} ,~~~~~~~A^{(0,f)}_{FB}=\frac{3}{4} A_e A_f,
\end{equation}
where, $g_V^f$ and $g_A^f$ are the effective vector and axial couplings of $Z$ to fermions ($f$ = $\tau$, $b$).

\begin{figure*}[htbp]
	\begin{subfigure}{0.5\textwidth}
		\centering
		\includegraphics[width=.95\linewidth]{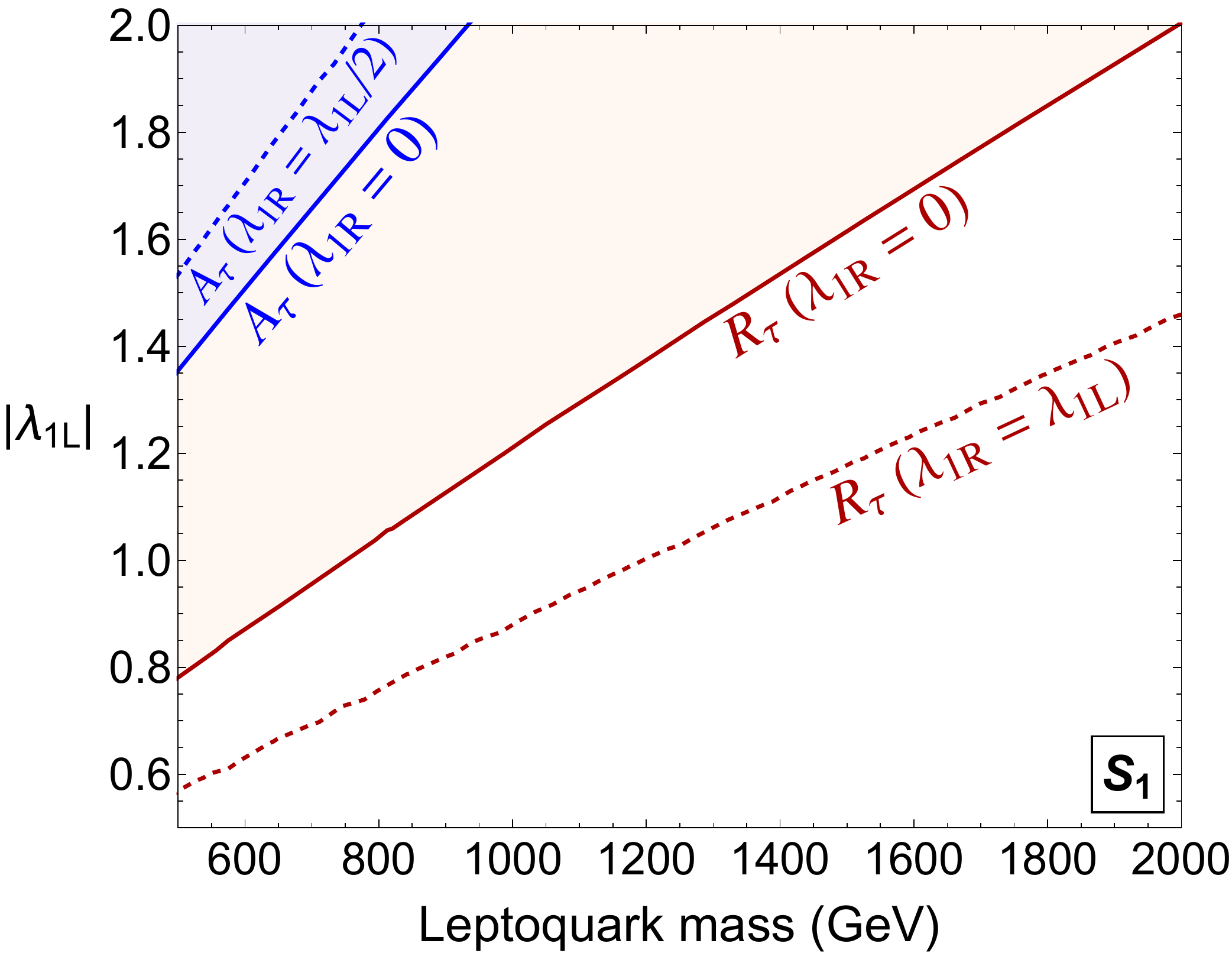}
	\end{subfigure}%
	\begin{subfigure}{0.5\textwidth}
		\centering
		\includegraphics[width=.97\linewidth]{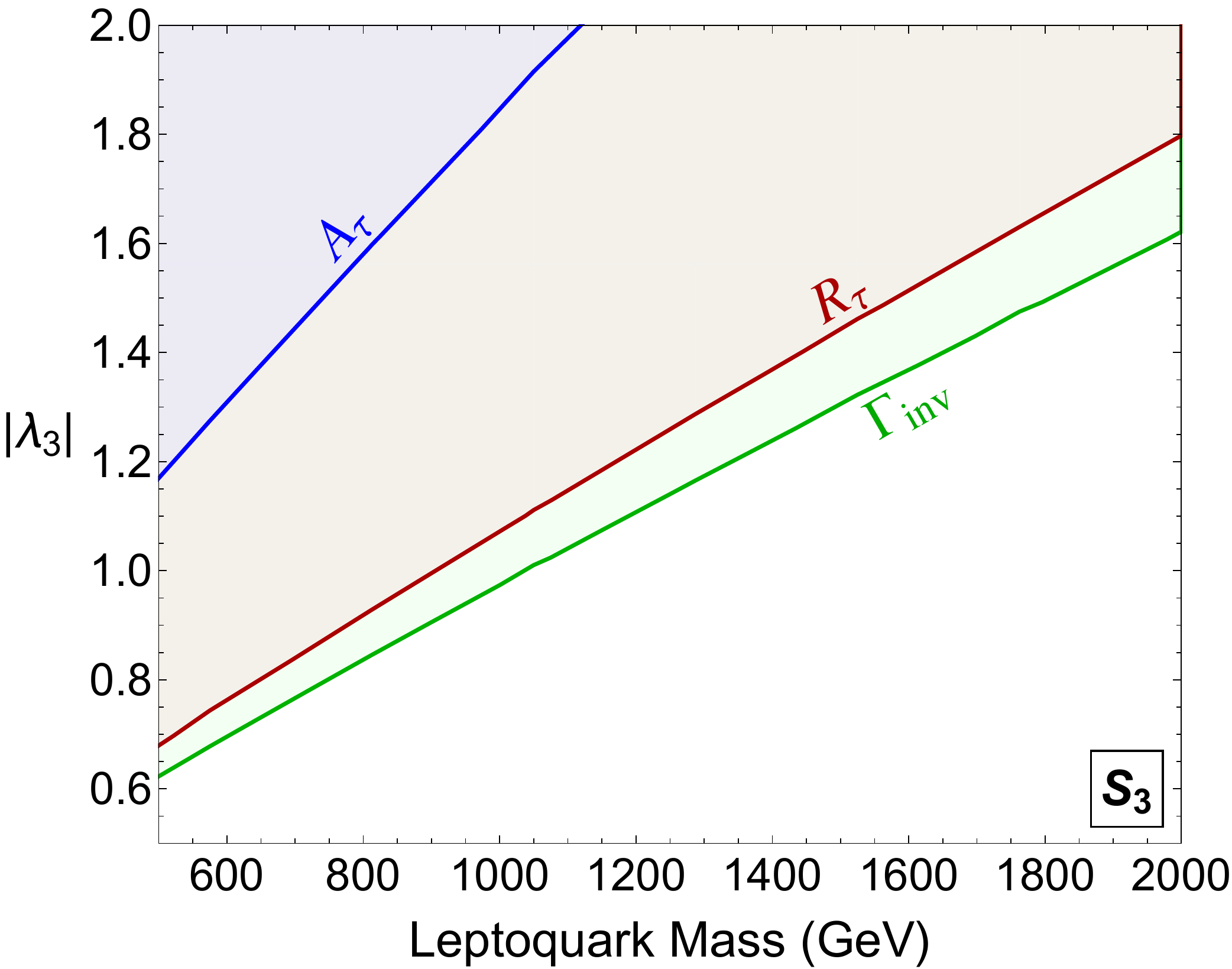}
	\end{subfigure}%
	\caption{\label{fig:S1LEP}Constraints from electroweak precision measurements on $S_1$(left) and $S_3$(right). Note that these constraints are based on $\chi^2$ rather than $\Delta\chi^2$. The lines indicate $\chi^2=5.99$ contours and the region above them is excluded by the corresponding observable.}
\end{figure*}

Figure~\ref{fig:S1LEP} summarizes the constraints from each of these electroweak measurements individually on $S_1$ and $S_3$.  The lines indicate contours of $\chi^2=5.99$ and the colored regions are disfavored by the corresponding observable. We consider the two leptoquarks now in turn.

\subsection{$\boldmath{S_1}$ Leptoquark}

The Lagrangian for $S_1$, Eq.~(\ref{AppLagS1}), along with Eqs.~(\ref{Appdeltagmz}) and (\ref{AppdeltagmL}), indicate that the correction to $Z \bar{\tau}  \tau$ coupling is directly proportional to $m_t^2$, whereas the corrections to $Z\bar{b} b$ and $Z\bar\nu_\tau \nu_\tau$ couplings are proportional to $m_Z^2$. Due to the $m_t$-enhancement, the most affected electroweak observables are the ones associated with the $Z\bar{\tau}\tau$ interaction. As we see in the left panel of Fig.~\ref{fig:S1LEP}, the strongest constraints do indeed arise from the measurement of $R_\tau$, followed by those from $A_\tau$. The effect of the leptoquark on $R_b$ and $A_b$ are too weak to impose any constraints in our region of interest, given their lack of $m_t$ enhancement; $\Gamma(inv)$ also provides no strong constraint due to its lack of $m_t$-enhancement and larger experimental uncertainty. It is also important to note that $S_1$ always gives a positive contribution to $R_\tau$ for all choices of couplings and masses. Because the SM prediction for $R_\tau$ is already somewhat higher than the experimental value, the $S_1$ leptoquark worsens the fit and is therefore more strongly constrained.

If we include the right-handed coupling (\ie, $\:\lambda_{1R} \neq 0$), the constraints from $R_\tau$  become even stronger, as shown by the dashed lines in the left panel of Fig.~\ref{fig:S1LEP}. On the other hand, constraints from the asymmetry parameter, $A_\tau$, are weakened, as the parity violation present in the model is diluted. $\Gamma (inv)$, $R_b$ and $A_b$ are not affected by $\lambda_{1R}$ as is evident from Eq.~(\ref{AppLagS1}). Because the overall constraints become stronger on including $\lambda_{1R}$, the bounds we obtained in the main text cannot be avoided by considering non-zero right-handed couplings.

\subsection{$\boldmath{S_3}$ Leptoquark}

In the triplet leptoquark case, both $Z\rightarrow \tau \bar{\tau}$ and $Z \rightarrow \nu_\tau \bar{\nu_\tau}$ get an $m_t$-enhancement, and thus observables associated with these decay channels are the most constraining. These observables include $R_\tau$, $A_\tau$ and $\Gamma(inv)$. In addition, the effect of $S_3$ on these observables is always positive, and thus, like $S_1$, $S_3$, also worsens the electroweak fit. These facts are reflected as strong constraints from these observables.

As shown in the right panel of Fig.~\ref{fig:S1LEP}, the invisible decay width of $Z$ imposes the strongest constraints, closely followed by $R_\tau$. Though both $R_\tau$ and $\Gamma(inv)$ get an $m_t$-dependent enhancement, the constraints from the latter are stronger because the SM prediction for $\Gamma(inv)$ is already higher than experimental value by $\sim1.7\sigma$ and the leptoquark-mediated process makes this discrepancy worse. As for the $S_1$ case, the observables associated with $Z\rightarrow b\bar{b}$ are barely affected by $S_3$ and do not show up in our plot.

It is important to note that the constraints obtained in this appendix were based on $\chi^2$, rather than $\Delta\chi^2$, for each observable individually. This allowed us to propertly account for the existing deviations between theory and the electroweak precision data. On the other hand, the electroweak constraints used in the main text are based on a combined $\Delta\chi^2 (=\chi^2-\chi^2_{SM})$ from all relevant observables.

\onecolumngrid

\end{document}